# Valley-dependent Multiple Quantum States and Topological Transitions in Germanene-based Ferromagnetic van der Waals Heterostructures

Feng Xue[1,2,*], Jiaheng Li[3], Yizhou Liu[4], Ruqian Wu[5], Yong Xu[2,6,7,*], Wenhui Duan[2,6,8]

[1] College of Physics & Optoelectronic Engineering, Jinan University, Guangzhou 510632, China

[2] State Key Laboratory of Low-Dimensional Quantum Physics, Department of Physics, Tsinghua University, Beijing 100084, China

[3] Beijing National Laboratory for Condensed Matter Physics and Institute of Physics, Chinese Academy of Sciences, Beijing 100190, China

[4] School of Physics Science and Engineering, Tongji University, Shanghai 200092, China

[5] Department of Physics and Astronomy, University of California, Irvine, CA 92697-4575, USA

[6] Frontier Science Center for Quantum Information, Beijing 100084, China

[7] RIKEN Center for Emergent Matter Science (CEMS), Wako, Saitama, 351-0198, Japan

[8] Institute for Advanced Study, Tsinghua University, Beijing 100084, China

## Abstract

Topological and valleytronic materials are promising for spintronic and quantum applications due to their unique properties. Using first principles calculations, we demonstrate that germanene (Ge)-based ferromagnetic heterostructures can exhibit multiple quantum states such as quantum anomalous Hall effect (QAHE) with Chern numbers of $C = -1$ or $C = -2$, quantum valley Hall effect (QVHE) with a valley Chern number of $C_v = 2$, valley-polarized quantum anomalous Hall effect (VP-QAHE) with two Chern numbers of $C = -1$ and $C_v = -1$ as well as time-reversal symmetry broken quantum spin Hall effect (*T*-broken QSHE) with a spin Chern number of $C_s \sim 1$. Furthermore, we find that the transitions between different quantum states can occur by changing the magnetic orientation of ferromagnetic layers through applying a magnetic field. Our discovery provides new routes and novel material platforms with a unique combination of diverse properties that make it well suitable for applications in electronics, spintronics and valley electronics.

\* Corresponding authors: fengxue@jnu.edu.cn; yongxu@mail.tsinghua.edu.cn;

## I. INTRODUCTION

Studies of topological materials have inaugurated a new era of discovery by unveiling quantum phenomena that intricately intertwine geometry, topology, and electronic states [1-4]. Among the plethora of quantum effects that have gained prominence, the quantum anomalous Hall effect (QAHE) is noteworthy for its potential to revolutionize modern electronic devices [5-7]. QAHE, an exotic state of matter, arises from the interplay between magnetism and topological order. It has been observed in magnetically doped topological insulators, such as Cr-doped $(Bi,Sb)_2Te_3$ [6]. In parallel, the quantum valley Hall effect (QVHE) has been identified as a phenomenon that highlights the unique valley physics present in 2D materials, showing potential for use in valleytronic applications [8-11]. Valleytronics seeks to control the energy extrema, or "valleys", within the electronic band structure, and adds a novel dimension of quantum functionality, supplementing charge and spin dynamics [12]. The coexistence of QAHE and QVHE may give rise to valley-polarized QAHE (VP-QAHE) [13,14], characterized by two nonzero Chern numbers of $C$ and $C_v$, representing the Chern number for QAHE and the valley Chern number for QVHE, respectively.

An emerging frontier in 2D materials research is the engineering of heterostructures, which allows for the combination of disparate materials to induce new or enhanced quantum phenomena [15,16]. As such, integrating topology, magnetism, and valley-dependent physics within a single system can give rise to multifunctional materials. In this regard, the fusion of germanene (Ge) with ferromagnetic (FM) materials, such as monolayers from the $MnBi_2Te_4$ (MBT) family, is especially intriguing. Germanene, a monolayer of germanium atoms configured in a buckled hexagonal honeycomb lattice [17,18], exhibits a Dirac cone near the Fermi level akin to graphene, yet it boasts the additional benefit of stronger spin-orbit coupling (SOC), which enhances its potential for spintronics and other related applications [19]. MBT, an intrinsic magnetic topological insulator recently uncovered [20], features an antiferromagnetic (AFM) van der Waals (vdW) layered structure. Depending on its thickness, MBT films may demonstrate alternating FM or AFM interlayer arrangements, giving rise to various topological phenomena, including QAHE, Weyl semimetal states, and axion insulator states. The coexistence of multiple magnetic and topological properties within MBT has sparked significant interest, leading to extensive theoretical [21-25] and experimental [26-29] investigations. The attractive properties of Ge, when allied with the intrinsic ferromagnetism of MBT family monolayers, set the stage for the exploration of diverse quantum phenomena.

In this work, through systematical first-principles calculations, we show that a variety of topological phases, including QAHE, QVHE, valley-polarized QAHE (VP-QAHE), and $T$-broken QSHE, can be realized in Ge-based FM vdW heterostructures, namely, Ge/$X$Bi$_2$Te$_4$ and $X$Bi$_2$Te$_4$/Ge/$Y$Bi$_2$Te$_4$ ($X$, $Y$ = V, Mn, Ni) heterostructures. The electrons around the Fermi level of these systems are 100% valley polarized with Dirac cone-like dispersion. We further find that by manipulating the magnetic orientation of the materials, topological phase transitions can be induced between different quantum states. For instance, flipping the spins from in-plane to out-of-plane can drive transitions from low-order QAHE ($C = \pm 1$) to high-order QAHE ($C = -2$), QVHE ($C_v = 2$) to QAHE ($C = -2$), and $T$-broken QSHE ($C_s \sim 1$) to QAHE ($C = -2$) or VP-QAHE ($C = -1, C_v = -1$), etc. Our findings not only provide a promising material platform for the realization of multiple quantum phenomena but also enable the induction of quantum phase transitions between different topological states through the application of external fields.

## II. COMPUTATIONAL METHODS

Density functional theory (DFT) calculations are performed using the Vienna *ab initio* simulation package (VASP) [30,31] within the generalized gradient approximation of the Perdew−Burke−Ernzerhof (GGA-PBE) [32] functional. The plane-wave cutoff energy is set to be 400 eV and a vacuum space larger than 15 Å is set to eliminate spurious interactions between periodic images. A Γ-centered 15 × 15 × 1 Monkhorst-Pack k-point mesh is used for structural relaxation, and a denser 23 × 23 × 1 mesh is used for magnetic anisotropy energies calculations. The in-plane lattice constants and atomic coordinates are fully relaxed until the force acting on each atom to be smaller than 0.005 eV/Å and energy converges to better than $10^{-6}$ eV. The strong electron correlation effects for the localized $d$ orbitals of V, Mn and Ni are treated by the DFT+$U$ method [33] using $U$ = 3, 4, 4 eV, respectively. The vdW corrections are invoked through the DFT-D3 method [34]. The maximally localized Wannier functions are constructed by using the software package Wannier90 [35] interfaced with VASP.

The Chern number and edge states are calculated by using Z2-Pack [36] and Wannier Tools [37].

## III. RESULTS

### A. Structural and magnetic properties

Here, three tetradymite-type magnetic materials, i.e., $X$Bi$_2$Te$_4$ ($X$BT with $X$ = Mn, V, Ni) consisting of a septuple layer (SL) of Te-Bi-Te-$X$-Te-Bi-Te, are considered. When combined with Ge, there are three types of Ge/$X$BT heterostructures and three types of sandwiched heterostructures $X$BT/Ge/$Y$BT ($X, Y$= Mn, V, Ni, $Y \neq X$), which exhibit spatial-inversion ($P$) symmetry breaking. Note that $Y \neq X$ is selected in the $X$BT/Ge/$Y$BT sandwich heterostructures in order to break the $P$ symmetry for bringing in valley-dependent electronic properties. There are six high-symmetry sites for Ge interfaced with $X$BT, depicted in Fig. S1. The calculated relative energies in Table S1 shows that the C1 configuration (Fig. S1a) is energetically most favorable. The structural details of this configuration are further shown in Fig. 1a, where the inner Ge atoms of Ge monolayer sit above the hollow sites of $X$BT. The fully relaxed lattice constants and interlayer distances for all the heterostructures are summarized in Table 1 and Table S1. Compared with those of Ge and $X$BT monolayers (Table S1), the lattice of Ge is stretched (~ 4%) and the counterparts of $X$BTs are slightly compressed (~ 1%) to form heterostructures.

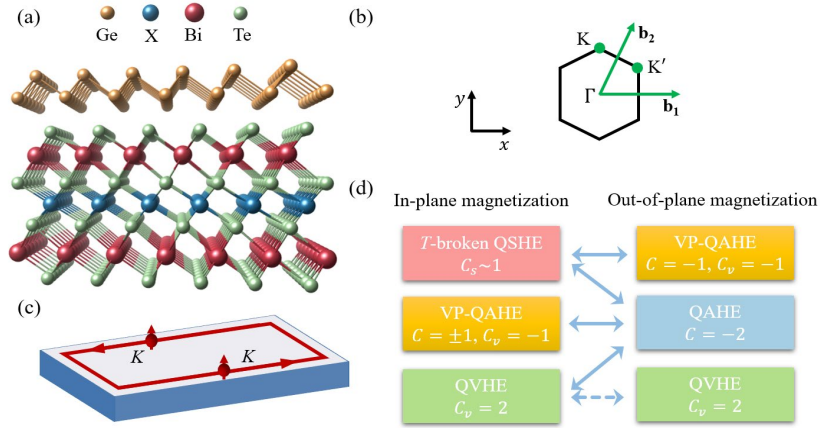

FIG. 1. (a) Crystal structure of $X$BT/Ge vdW heterostructures. (b) The first Brillouin zone (BZ) of $X$BT/Ge and $X$BT/Ge/$Y$BT heterostructures, where $K$ and $K'$ points are marked by green dots. (c) Schematic plot of VP-QAHE with one valley-dependent chiral edge state. (d) Schematic diagram of topological transitions in Ge-based ferromagnetic heterostructures without $P$, where solid double-headed arrows represent the transitions between two different quantum states, and dashed double-headed arrows indicate that the system's quantum states remain unchanged during the magnetization reversal.

The magnetic parameters, including exchange parameters J, magnetocrystalline anisotropy (MCA), total magnetic moments ($M_{tot}$), are calculated for the ground state geometries (Table I). Exchange interaction parameters are obtained by fitting the total energies from DFT calculations of various spin configurations (Fig. S4) to the Heisenberg spin Hamiltonian, $H = -J\sum_{<ij>} S_i \cdot S_j$, with interactions among the first nearest intralayer and interlayer neighbors only. As shown in Table 1, all systems prefer the ferromagnetic order except for MBT/Ge/NBT which has an intralayer FM coupling and interlayer AFM coupling. The MCA is calculated by taking energy difference as MCA = $E_\parallel - E_\perp$, where $E_\parallel$ and $E_\perp$ represent energies of in-plane and out-of-plane magnetization directions, respectively. It can be seen that the systems of VBT/Ge and $X$BT/Ge/VBT that contains VBT always have negative values of MCA, indicating in-plane magnetic easy axis. Apart from them, other systems prefer out-of-plane magnetization. As magnetic field can tune the spin orientation toward a specific direction, all calculations in the following are based on the out-of-plane ferromagnetic configuration unless indicated otherwise.

Table 1. Properties of considered systems, including calculated interlayer distance $d$ (Å), exchange parameters $J$ (meV), magnetocrystalline anisotropy MCA (meV), total magnetic moment $M_{tot}$ ($\mu_B$), band gap at $K$ $\Delta_K$ and $K'$ points $\Delta_{K'}$ (meV). The values on the left and right sides of the slash correspond to the parameters of left and right parts of the heterostructures, respectively. Particularly, the middle value between two slashes of $J$ refer to interlayer exchange parameter.

| Systems | $d$ | $J$ | MCA | $M_{tot}$ | $\Delta_K$ | $\Delta_{K'}$ |
|---|---|---|---|---|---|---|
| Ge/VBT | 2.81 | 3.03 | -0.22 | 3.0 | 70 | 51 |
| Ge/MBT | 2.77 | 1.26 | 0.26 | 5.0 | 14 | 74 |
| Ge/NBT | 2.75 | 0.93 | 0.04 | 2.0 | 142 | 106 |
| MBT/Ge/VBT | 2.64/2.61 | 1.05/0.03/4.46 | -0.17 | 8.0 | 29 | 62 |
| MBT/Ge/NBT | 2.59/2.56 | 0.94/-0.20/1.40 | 0.18 | 3.0 | 181 | 229 |
| NBT/Ge/VBT | 2.57/2.64 | 1.52/0.42/4.14 | -0.32 | 5.0 | 59 | 174 |

### B. Topological properties

Compared to graphene, germanene possesses the buckled structure, with two sublattices displaced vertically. Its moderate spin-orbit coupling strength opens a direct gap for the Dirac bands at the $K/K'$ points (Fig. S3). It has been recognized that germanene belongs to $T$ invariant topological insulators with $Z_2 = 1$. For heterostructures composed of germanene and XBT, both $T$ and $P$ symmetries are broken, and its original topology ($Z_2$) cannot be retained, but new magnetic topological phases emerge. According to previous theoretical work [38], germanene can exhibit various topological states when $T$ and $P$ symmetries are broken, and these states can be classified into four main categories: (1) QVHE with a nonzero valley Chern number of $C_v$, (2) QAHE with nonzero Chern number of $C$, (3) VP-QAHE with two nonzero Chern numbers of $C$ and $C_v$, (4) $T$-broken QSHE with a nonzero valley Chern number of $C_s$. Since there have been numerous studies on low-energy Hamiltonian models based on germanene [38] or silicene [38,39], we will not delve into this issue here. Instead, our work primarily focuses on identifying practical materials to realize these topological states.

The topological numbers of Ge-based vdW heterostructures are listed in Table 2. To illustrate the physics, Figure 2 presents three representative topological phases, i.e., the QVHE, QAHE, and VP-QAHE phases. All three systems appear to have similar band structures where bands around the Fermi level are 100% valley-polarized with Dirac cone-like dispersion, but they have distinct features in detail. Figure 2a displays the band structures of VBT/Ge heterostructure along high symmetry path (Fig. 1b), weighted with VBT- and Ge-projection (top panel). Obviously, states near the Fermi level are mainly contributed by the $p_z$ orbitals of Ge (Fig. S4), with valence band maximum (VBM) and conduction band minimum (CBM) located at $K$ and $K'$ points. In contrast to pristine Ge, which exhibits a band gap of 24 meV at both $K$ and $K'$ points (Fig. S3b), VBT/Ge displays different values, 70 meV and 51 meV, at the $K$ and $K'$ points, respectively. To confirm that QVHE can indeed be realized in this system, we calculated the distribution of Berry curvature ($\Omega$) in the first Brillouin zone (BZ) (Fig. 2c). The Berry curvature dominantly distributes in the vicinities of $K$ and $K'$ valleys and with opposite signs around those points, reflecting the physics of coupled spin and valley. Furthermore, the Berry curvature along the high-symmetry lines is shown in Fig. 2a, where two peaks with opposite values at $K$ and $K'$ points can be seen clearly. After integration of the Berry curvatures over the whole first BZ, we obtain Chern number $C = 0$. However, separate integrations in the triangular regions centered at $K$ and $K'$ valleys give two values of $C_K = 1$ and $C_{K'} = -1$, which corresponds to $C_v = 2$, indicating features of the topological QVH phase with $(C, C_v) = (0, 2)$.

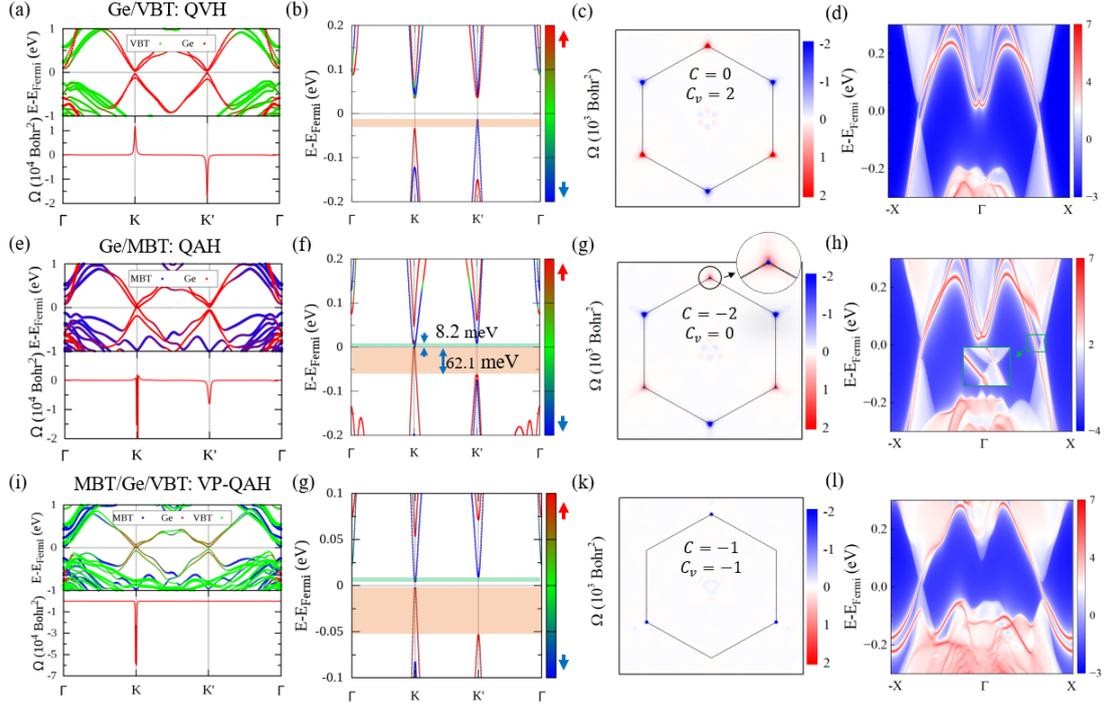

FIG. 2. (a) VBT- and Ge-projected band structures with SOC included (top panel), and the Berry curvature distribution along high symmetry path (bottom panel). (b) The spin-projected (out of plane $<s_z>$) band structure of VBT/Ge with SOC included. The red and blue colors denote the spin up and spin down projection. (c) The distribution of Berry curvature ($\Omega$) of VBT/Ge in the first BZ. (d) The calculated nontrivial chiral edge states. (e)-(h) and (i)-(l) are same as (a)-(d) but for Ge/MBT and MBT/Ge/VBT heterostructures, respectively. The light yellow and green shadows in (b), (f) and (g) indicate the valley splitting between $K$ and $K'$ valleys.

To obtain the QAHE, a stronger exchange field is needed [39]. This indicates that MBT/Ge or $X$BT/Ge/$Y$BT are more suitable for this purpose as they have stronger magnetic moment compared to VBT/Ge. To verify the hypothesis, we calculate the electronic and topological properties of MBT/Ge and other systems. As illustrated in Figs. 2e-f, the component-projected band structure of MBT/Ge is overall similar to that of VBT/Ge, but distinct behaviors are evident near the Fermi level in the $K$ and $K'$ valleys. At the $K$ point, the energy gap reemerges after a band crossing but is narrowed to 2 meV. Conversely, the energy gap at the $K'$ point is widened to 73 meV. Importantly, the band inversion at the $K$ point leads to the change of sign of the Berry curvature, from positive in VBT to negative in MBT at the $K$ point. In contrast, the sign at the $K'$ point remains unchanged. This observation is evident from the distribution of Berry curvature across the BZ shown in Fig. 2g and along the high-symmetry path presented in Fig. 2e. The integration of Berry curvature around the $K$ and $K'$ points yields valley

Chern numbers $C_K = -1$ and $C_{K'} = -1$, respectively, further confirming this behavior. Integrating over the entire BZ results in a total Chern number of $C = -2$, and the observation of two chiral edge states for an MBT/Ge ribbon, as depicted in Fig. 2h, collectively confirms the manifestation of the QAHE. The same topological phase can also be obtained for NBT/Ge, in which the nontrivial band gap is enhanced to 19 meV (Fig. S5). In addition, other two systems of MBT/Ge/NBT and NBT/Ge/VBT also exhibit valley-dependent QAHE with $C = -2$, and the detail electronic and topological properties are shown in Fig. S6. Note that, the phases with opposite Chern number $C$ can be obtained by reversing the magnetization.

Now we investigate the MBT/Ge/VBT heterostructure to determine if it is possible that both QAHE and QVHE coexist in a single system. Figure 2i shows the component-projected band structure of MBT/Ge/VBT, in which blue, red and green solid circles represent MBT, Ge and VBT components, respectively. Interestingly, bands near the Fermi level at the K' valley contain contributions from both Ge and VBT, different from aforementioned cases for which Ge $p_z$ orbitals dominate. The enhanced hybridization between the Ge and VBT in the trilayer sandwich is due to the reduced interlayer distance comparing with the bilayer ones (Table I). As a result, the Berry curvature nullifies at $K'$ point but remains at $K$ point as shown in Figs. 2i and 2j. By integrating Berry curvature throughout the BZ, the Chern number $C = -1$, with $C_K = -1$ and $C_{K'} = 0$, is obtained. That means the QAHE is valley polarized with $C_v = -1$, which only occurs near $K$ point as shown in Figs. 2i and 2k. For the VP-QAHE, the edge states exhibit chiral-spin-valley locking effect as show in Figs. 2l and 1c, where one chiral edge state only appears at $K$ valley with spin-polarized carriers.

It is intriguing that these systems not only display distinct topological states but also possess excellent valley electronic properties. In the absence of SOC, the spin up and spin down channels split at $K$ and $K'$ valleys due to the $T$ and $P$ symmetry breaking (Fig. S5a). Upon including SOC, the band degeneracy at the $K$ and $K'$ valleys is lifted and the spontaneous valley splitting occurs, as shown in Figs. 2b, 2f and 2j. To quantitatively describe the scales of spin splitting and valley splitting, we define the spin splitting [40] as $\Delta_{v(c)}^{\tau} = E_{\uparrow}^{v(c),\tau} - E_{\downarrow}^{v(c),\tau}$ and valley splitting [41] as $\Delta_{KK'}^{v(c)} = |E_K^{v(c)} - E_{K'}^{v(c)}|$, where $v(c)$ represents valence (conduction) bands and $\tau$ denotes the index of $K$ and $K'$ valleys. The former indicates the spin splitting of valence (conduction) bands at $K$ and $K'$ points, while the latter gives the energy difference between the top (bottom) of valence (conduction) bands at the $K$ and $K'$ valleys. According to this definition, the calculated spin splitting of $\Delta_v^K$, $\Delta_v^{K'}$, $\Delta_c^K$, $\Delta_c^{K'}$ and valley splitting of $\Delta_{KK'}^v$, $\Delta_{KK'}^c$ are listed in Table S3. It is worth to note that the spin splitting in all systems is sizable and up to 196 meV in MBT/Ge. What is more, the $\Delta_{KK'}^v$ in Ge/MBT is up to 62 meV, corresponding to the valley splitting generated by a magnetic field of 620 T. Therefore, all systems exhibit spontaneous valley splitting and chiral edge states (except for VBT/Ge), and can be considered as excellent candidates for designing dissipationless valleytronics.

Table 2. Topological states (TS) for different magnetization directions.

| Systems | Out-of-plane magnetization | | | | | | In-plane magnetization | | | | | |
|---|---|---|---|---|---|---|---|---|---|---|---|---|
| | $C_K$ | $C_{K'}$ | $C$ | $C_v$ | $C_s$ | TS | $C_K$ | $C_{K'}$ | $C$ | $C_v$ | $C_s$ | TS |
| VBT/Ge | 1 | -1 | 0 | 2 | 0 | QVH | 1 | -1 | 0 | 2 | 0 | QVH |
| MBT/Ge | -1 | -1 | -2 | 0 | 0 | QAH | 1 | -1 | 0 | 2 | 0 | QVH |
| NBT/Ge | -1 | -1 | -2 | 0 | 0 | QAH | 0/-1 | 1/0 | ±1 | -1 | 1 | VP-QAH |
| MBT/Ge/VBT | -1 | 0 | -1 | -1 | 1 | VP-QAH | 0 | 0 | 0 | 0 | 1 | *T*-broken QSH |
| MBT/Ge/NBT | -1 | -1 | -2 | 0 | 0 | QAH | 0 | 0 | 0 | 0 | 1 | *T*-broken QSH |
| NBT/Ge/VBT | -1 | -1 | -2 | 0 | 0 | QAH | 0 | 0 | 0 | 0 | 1 | *T*-broken QSH |

### C. Magnetic orientation effects

As we mentioned above, the systems including VBT tend to form in-plane magnetization, so it is necessary to explore the effects of magnetic orientation. As summarized in Table II, there are 3 topological states when in-plane magnetization is assumed: (1) for $X$BT/Ge ($X$ = Mn, V) bilayer heterostructures, QVHE as in VBT/Ge exists. (2) for NBT/Ge, VP-QAHE with $C = \pm 1$ is induced, where tunable QAHE [25,42] can be obtained by tuning the magnetic orientation. (3) for $X$BT/Ge/$Y$BT heterostructures, *T*-broken QSHE presences. The details regarding the former two phase are presented in supplemental materials, we now focus on the last one: the so-called *T*-broken QSHE [43].

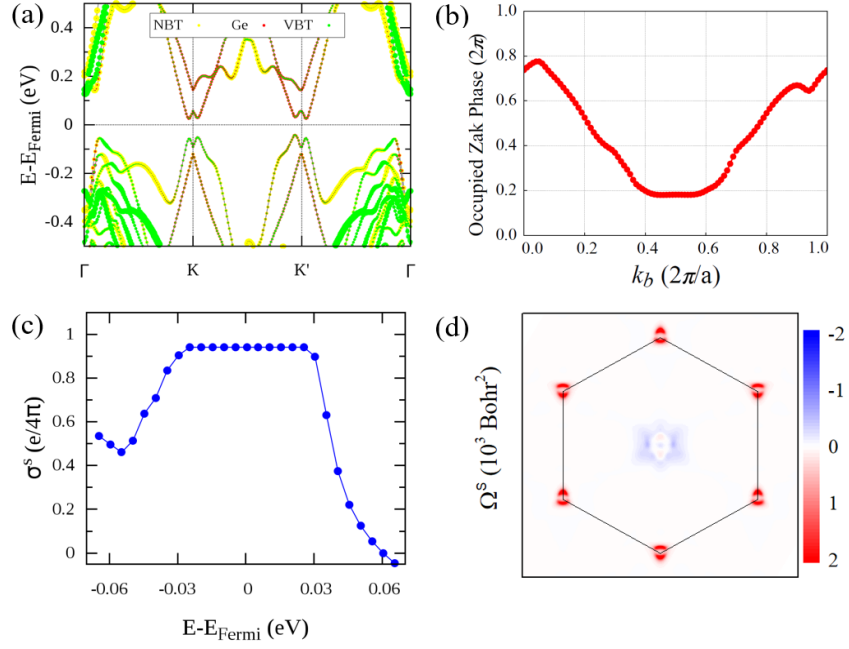

FIG. 3. The *T*-broken QSHE in the sandwiched NBT/Ge/VBT heterostructure for in-plane magnetization. (a) NBT-, Ge- and VBT-projected band structure with SOC included. The yellow, red and green solid circles, symbolled by their weights, represent the NBT, Ge and VBT components, respectively. (b) The evolution of WCC in the full closed plane. (c) The spin Hall conductance $\sigma^s$ as a function of the Fermi level. (d) Distribution of spin Berry curvature in first BZ.

Taking VBT/Ge/NBT as an example, Fig. 3a displays the NBT-, Ge- and VBT-projected band structure. Clearly, states near the Fermi level at $K$ and $K'$ valleys mainly come from Ge and VBT, implying that the Berry curvature may disappear at two valleys. To verify our conjecture, the Berry curvature distribution (not shown) and the evolution of Wannier charge centers (WCC) for the full closed plane (Fig. 3b) is calculated using the Wilson loop method, indicating that this system with in-plane magnetization is not a QAH insulator ($C = 0$). However, the calculated spin Berry curvature [44,45] in the first BZ reveals a large positive value in vicinities of $K$ and $K'$ valleys, but a negative value near $\Gamma$ point (Fig. 3d). By integrating the spin Berry curvature over the whole BZ, we obtain the spin Chern number $C_s \sim 1$. Actually, integrations in the triangular regions centered at $\Gamma$ point as well as $K$ and $K'$ valleys whose area is equal to half of the first BZ give $C_s^\Gamma \sim -1$, $C_s^K \sim 1$ and $C_s^{K'} \sim 1$. The spin Hall conductance $\sigma^s$ as a function of the Fermi level ($E_F$) is given in Fig. 3c, where a clear plateau of $\sigma^s = C_s \frac{e}{4\pi}$ with $C_s \sim 1$ can be observed. Thus, VBT/Ge/NBT system with in-plane magnetization is a *T*-broken QSH insulator.

## IV. DISCUSSION AND CONCLUSION

Experimentally, magnetic fields are often used to change the magnetization orientation or to induce magnetic ordering. In our systems, if an external magnetic field is employed to turn the magnetization direction, for example, from in-plane to out-of-plane, the transitions between two different quantum states can be observed for all systems except VBT/Ge for which the QVHE is persistent, independent of the magnetization direction. As summarized in Table II, one may see that the transitions of MBT/Ge from QVH to QAH, NBT/Ge from VP-QAH with $C = \pm 1$ to QAH with $C = -2$, MBT/Ge/VBT from $T$-broken QSH to VP-QAH states, can be achieved. Possible topological phase diagrams as the magnetization direction varies are further illustrated in Fig. 1d. It is worth noting that $X$BT/Ge/$Y$BT heterostructures can still possess nontrivial topological properties even $X$BT and $Y$BT has AFM coupling, due to their noncompensated magnetic moments and proximity effects. Therefore, richer topological transitions may occur in these systems. Some discussions in those aspects are provided in the supplementary materials. For XBT/Ge/XBT with $P$ symmetry, only $T$-broken QSHE and QAHE can be realized (Fig. S10). Considering that MBT [46] and Ge [47,48] thin films have been successfully synthesized, we may expect that the Ge-based heterostructures studied in this work can be fabricated experimentally.

In summary, first-principles calculations have been performed to investigate the structural, electronic, magnetic and topological properties of heterostructures constructing with Ge and $X$BT ($X$ = V, Mn, Ni) monolayers. We show that the systems can achieve multiple valley-dependent quantum states, i.e., QAHE, $T$-broken QSHE, QVHE, and VP-QAHE. More importantly, diverse topological quantum transitions, such as from QVHE to QAHE, can be controlled by switching magnetization between in-plane and out-of-plane directions with magnetic field, strain or even electric bias. Furthermore, the significant splittings in the valleys, with values as high as 196 meV and 62 meV for the spin splitting and valley splitting, hold tremendous promise for applications in dissipationless spintronic and valleytronic devices. Our findings offer various material platforms for realizing multiple quantum states that are essential for technological developments.

## ACKNOWLEDGMENTS

F. X. would like to thank Zhe Wang, Tong Zhou and Yusheng Hou for helpful discussions. This work was supported by the Ministry of Science and Technology of China (Grant No. 2023YFA1406400), the Basic Science Center Project of NSFC (Grant No. 51788104), the National Science Fund for Distinguished Young Scholars (Grant No. 12025405), the Beijing Advanced Innovation Center for Future Chip, the Beijing Advanced Innovation Center for Materials Genome Engineering, and China Postdoctoral Science Foundation (grant No. 2023M741388). Work at the University of California, Irvine was supported by DOE-BES of USA (Grant No. DE-FG02-05ER46237). Computer simulations were performed at the U.S. Department of Energy Supercomputer Facility (NERSC) and at Hefei advanced computing center.